\documentclass[11pt]{article}
\usepackage{moriond,epsfig}

\def\gev{\, \mbox{GeV}}

\def\KL{K\"all\'en-Lehmann}
\def\Journal#1#2#3#4{{#1} {\bf #2}, #3 (#4)}

\def\PLB{{\em Phys. Lett.}  B}
\def\PRL{\em Phys. Rev. Lett.}
\def\PRD{{\em Phys. Rev.} D}

\def\be{\begin{equation}}
\def\ee{\end{equation}}
\def\bea{\begin{eqnarray}}
\def\eea{\end{eqnarray}}

\begin{document}
\vspace*{4cm}
\title{NO HIGGS AT THE LHC}

\author{ J. J. VAN DER BIJ }

\address{Institut f\"ur Physik, Albert-Ludwigs Universit\"at Freiburg, H. Herderstr. 3,\\
79104 Freiburg i.B., Deutschland}

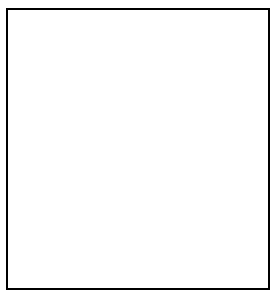
\maketitle\abstracts{I discuss the question whether it is possible that the LHC
will find no signal for the Higgs particle. It is argued that in this case singlet scalars
should be present that could move in extra, even fractional, dimensions.
A critical view at the existing electroweak data shows that this possibility
might be favored over the simplest standard model. In this case one needs the ILC
in order to study the Higgs sector.
}

\section{Introduction}
The standard model gives a good description of the bulk of the electroweak data.
Only a sign of the Higgs particle is  missing at the moment. The Higgs field is
necessary in order to make the theory renormalizable, so that predictions are
possible and one can really speak of a theory. A complete absence of the 
Higgs field would make the theory non-renormalizable, implying the existence
of new strong interactions at the TeV scale. Therefore one is naively led to the
so-called no-lose theorem~\cite{chanowitz}. This theorem says that when one
 builds a large energy
hadron collider, formerly the SSC now the LHC, one will find new phyics, either 
the Higgs particle or otherwise new strong interactions. Since historically
no-theorems have a bad record in physics one is naturally tempted to try
to evade this theorem. So in the following I will try to find ways by which
the LHC can avoid seeing any sign of new physics. 

At the time of the introduction of the no-lose theorem very little was known
about the Higgs particle. Since then there have been experiments at LEP, SLAC and
the Tevatron, that give information on the Higgs mass. Through precise measurements
of the W-boson mass and various asymmetries one can get constraints on the
Higgs mass. The Higgs mass enters into the prediction of these quantities via radiative
corrections containing a virtual Higgs exchange. Moreover at LEP-200 the direct search
gives  a  lower limit of $114.4 \gev$. The situation regarding the precision tests 
is not fully satisfactory. The reason is that the Higgs mass implied by the 
forward-backward asymmetry $A_{FB}(b)$ from the bottom quarks is far away from the
mass implied by the other measurements, that agree very well with each other.
No model of new physics appears to be able to explain the difference.
From $A_{FB}(b)$ one finds $m_H= 488^{+426}_{-219} \gev$ with a $95\%$ lower bound
of $m_H=181 \gev$. Combining the other experiments one finds
$m_H= 51^{+37}_{-22} \gev$ with a $95\%$ upper bound of $m_H=109 \gev$. The $\chi^2$
of the latter fit is essentially zero. Combining all measurements gives a bad fit.
One therefore has a dilemma. Keeping all data one has a bad fit. Ignoring the 
$b$-data the standard model is ruled out. In the latter case one is largely forced 
towards the extended models that appear in the following. Accepting a bad fit  one has
somewhat more leeway, but the extended models are still a distinct possibility.

\section{Hiding the Higgs boson}
\subsection{Invisible decay}
When singlet scalar fields are added to the Higgs sector two effects appear,
invisible decay and mixing. The general class of models is described in ref.~\cite{bij2006}.
In the case that there is an unbroken symmetry in the singlet sector the singlets
are stable and weakly interacting. They are a perfectly simple candidate for
the dark matter of the universe. If they are light enough the Higgs boson will decay 
 into these invisible particles, thereby leaving no obvious signal at the LHC.

However, is this Higgs boson completely undetectable at the LHC?
Its production mechanisms are exactly the same as the standard model 
ones, only its decay is in undetectable particles. One therefore
has to study associated production with an extra Z-boson or one must
consider the vector-boson fusion channel with jet-tagging. Assuming
the invisible branching ratio to be large and assuming the Higgs boson
not to be heavy, as indicated by the precision tests, one still finds
a significant excess of events. Of course one cannot study
this Higgs boson in great detail at the LHC. For this the ILC
would be needed, where precise measurements are possible in the
channel $e^+e^-\rightarrow Z H$.

\subsection{Mixing: fractional Higgses}

Somewhat surprisingly it is possible to have a model
that has basically only singlet-doublet mixing. If one starts with an interaction of the form
$H \Phi^{\dagger}\Phi$, where H is the new singlet Higgs field and $\Phi$ the standard model
Higgs field, no interaction of the form $H^3$, $H^4$ 
or $H^2 \Phi^{\dagger}\Phi$ is generated with an infinite
coefficient~\cite{hill}. 
At the same time the scalar potential stays bounded from below.
This means that one can indeed leave these dimension four interactions out of the Lagrangian
without violating renormalizability. This is similar to the non-renormalization theorem
in supersymmetry that says that the superpotential does not get renormalized.
However in general it only works with singlet extensions. As far as the counting of parameters
is concerned this is the most minimal extension of the standard model, having
only two extra parameters.

 The simplest model is the Hill model: 
\begin{equation}
L = -\frac{1}{2}(D_{\mu} \Phi)^{\dagger}(D_{\mu} \Phi) 
-\frac {1}{2}(\partial_{\mu} H)^2 - \frac {\lambda_0}{8}
(\Phi^{\dagger} \Phi -f_0^2)^2  -
\frac {\lambda_1}{8}(2f_1 H-\Phi^{\dagger}\Phi)^2 
\end{equation}
Working in the unitary gauge one writes $\Phi^{\dagger}=(\sigma,0)$,
where the $\sigma$-field is the physical standard model Higgs field.
Both the standard model Higgs field $\sigma$ and the Hill field $H$ receive vacuum expectation
values and one ends up with 
 two particles having the quantum numbers of the Higgs particle,
but reduced couplings to standard model particles. One can call them fractional Higgs
particles.
A practical way to describe the situation is to replace in all experimental cross section
calculations the standard model Higgs propagator by:
\begin{equation}
D_{\sigma\sigma} (k^2) = \cos^2(\alpha)/(k^2 + m_-^2) + \sin^2(\alpha)/(k^2 + m_+^2)
\end{equation}
The generalization to an arbitrary set of fields $H_i$ is straightforward.
One finds a number of (fractional) Higgs bosons $H_i$ with reduced
couplings $g_i$ to the standard model particles.

\subsection{A higher dimensional Higgs boson}
The mechanism described above can be generalized to an infinite number
of Higgses. The physical Higgs propagator is then given by an infinite number
of very small Higgs peaks, that cannot be resolved by the detector.
Ultimately one can take a continuum limit,
so as to produce an arbitray line shape for the Higgs boson, satisfying
the K\"all\'en-Lehmann representation.

\be D_{\sigma \sigma}(k^2)= \int ds\, \rho(s)/(k^2 + \rho(s) -i\epsilon) \ee

One has the sum rule~\cite{akhoury,gunion}
 $\int \rho(s)\, ds = 1$, while otherwise the theory is not renormalizable
and would lead to infinite effects for instance on the LEP precision variables.
Moreover, combining mixing with invisible
decay,  one can vary the invisible decay branching ratio
as a function of the invariant mass inside the Higgs propagator.
There is then no Higgs peak to be found any more.
The general Higgs propagator for the Higgs boson in the presence of
singlet fields is therefore determined by two functions, the \KL\, spectral density
and the s-dependent invisible branching ratio. Unchanged compared to the
standard model are the relative branching ratio's to standard model particles.

Since a sharp mass peak is absent, this is a promising way to hide the Higgs boson at the LHC.
The general case is rather arbitrary, but it contains
an elegant subclass. Because the $H \Phi^{\dagger}\Phi$ interaction
is superrenormalizable one can  let the $H$ field move in more
dimensions than four, without violating renormalizability. 
One can go up to six dimensions. The precise form of the propagator
depends on the size and shape of the higher dimensions.
The exact formulas can be quite complicated. However it is possible that
these higher dimensions are simply open and flat. In this case one finds
simple formulas. One has for the generic case a propagator of the form:

\be
D_{\sigma \sigma}(q^2)= \left[ q^2 +M^2 - \mu_{lhd}^{8-d}
(q^2+m^2)^{d-6 \over 2} \right]^{-1} .\ee

%

The parameter $M$ is a four-dimensional mass, $m$ a higher-dimensional mass
and $\mu_{lhd}$ a higher-to-lower dimensional mixing mass scale.
For six dimensions a limiting procedure is needed.
Calculating the corresponding \KL\, spectral densities one finds
a low mass peak and a continuum that starts a bit higher in the mass.



If one does not introduce further fields no invisible decay is present.
If the delta peak is small enough it is too insignificant for the LHC search.
The continuum is practically invisible at the LHC, since the low mass Higgs search 
depends on the presence of a sharp peak.

\section{Comparison with the LEP-200 data}
We now confront the higher dimensional models with the results from the direct
Higgs search at LEP-200~\cite{lep200}.
Within the pure standard model the absence of a clear signal has led to
a lower limit on the Higgs boson mass of $114.4 \gev$ at the 95\% confidence level.
Although no clear signal was found the data have some intriguing features,
that can be interpreted as evidence for Higgs bosons beyond the standard
model. There is a $2.3\,\sigma$ effect seen by all experiments at around 98 GeV.
A somewhat less significant $1.7\,\sigma$ excess is seen around 115 GeV. Finally
over the whole range $s^{1/2} > 100\gev$ the confidence level is less than
expected from background.
We interpet these features as evidence for a spread-out Higgs-boson~\cite{dilcher}.
The peak at $98 \gev$ is taken to correspond to the delta peak in the
\KL\, density. The other excess data are interpreted as part of the continuum,
that peaks around $115 \gev$. We  vary the data within the uncertainties
of the experiment. 
We start with the case $d=5$.
There is no problem fitting the data with these conditions. As allowed ranges
we find:
\begin{eqnarray}
& 95\gev< m < 101\gev \nonumber\\
& 111\gev< M < 121\gev \nonumber\\
& 26\gev < \mu_{lhd} < 49\gev 
\end{eqnarray}

We now repeat the analysis for the case $d=6$. Here the data can only be fitted in
a restricted range within the uncertainties of the experiment. 
The six-dimensional
case therefore seems to be somewhat disfavoured compared to the five-dimensional
case. 
We found the following limits:

\begin{eqnarray}
& 95\gev< m < 101\gev \nonumber\\
& 106\gev < M < 111\gev \nonumber\\
& 22\gev < \mu_{lhd} < 27\gev 
\end{eqnarray}

\section{Conclusion}
Taking the analysis at face value we find a roughly $3 \sigma$ effect
with the following conclusions:\\

\noindent a) The Higgs field has been found at LEP-200.\\
b) Its properties are consistent with the electroweak precision data.\\
c) A dark matter candidate can be included.\\
d) The LHC will see no Higgs signal.\\

We note that also the Tevatron appears to have an excess at low masses in the
Higgs search. However the experimental significance is hard to estimate, as the
data were not analysed with this type of model in mind.

\section*{Acknowledgments}
This work was supported by the BmBF Schwerpunktsprogramm
"Struktur und Wechselwirkung fundamentaler Teilchen".

\end{document}